# Detecting changes in anthropogenic light emissions: limits due to atmospheric variability.

Salvador Bará

*Independent scholar. Former profesor titular (retired) at Universidade de Santiago de Compostela.*
*15703 Santiago de Compostela, Galicia (Spain, European Union)*

e-mail: salva.bara@usc.gal – https://orcid.org/0000-0003-1274-8043

**Abstract:** Monitoring the evolution of the anthropogenic light emissions is a priority task in light pollution research. Among the complementary approaches that can be adopted to achieve this goal stand out those based on measuring the direct radiance of the sources at ground level or from low Earth orbit satellites, and on measuring the scattered radiance (known as artificial night sky brightness or skyglow) using networks of ground-based sensors. The terrestrial atmosphere is a variable medium interposed between the sources and the measuring instruments, and the fluctuation of its optical parameters sets a lower limit for the actual source emission changes that can be confidently detected. In this paper we analyze the effect of the fluctuations of the molecular and aerosol optical depths. It is shown that for reliably detecting changes in the anthropogenic light emissions of order ~1% per year, the inter-annual variability of the annual means of these atmospheric parameters in the measurement datasets must be carefully controlled or efficiently corrected for.

## 1. Introduction

The development of instruments and methods for detecting changes in the anthropogenic light emissions is a pressing issue in light pollution research. Broadly speaking, there are three main complementary approaches that may be applied to this problem: (i) the administrative one, based on the analysis of updated inventories of existing outdoor lights, (ii) direct radiance sensing, whereby the direct light from the sources is measured by radiometers pointing at them from nearby or distant places, and (iii) scattered radiance sensing, based on measuring the anthropogenic night sky brightness or skyglow, that is, the artificial light scattered by the atmosphere.

Direct radiance sensing has a long tradition in light pollution studies. Remote sensing of nighttime lights using radiometers onboard low Earth orbit satellites (Elvidge et al. 2022; Li et al. 2017; Román et al. 2018) and nighttime images from the International Space Station (ISS) (Kyba et al. 2015; Sánchez de Miguel et al. 2019; Stefanov et al. 2017) provide very valuable information about the global extent of light pollution (Falchi et al. 2016), the rate of increase of the total light emissions (Kyba et al. 2017), and the changes in their spectral composition (Sánchez de Miguel et al. 2021, 2022). Working at ground level and at shorter distances from the sources, direct measurements of outdoor lights have been successfully demonstrated for a variety of purposes by "urban observatories" (Dobler et al. 2015, 2016, 2021). The instruments and methods of these observatories, e.g. time-lapse imaging of city lights from a vantage site, can also be applied to the study of the long-term evolution of the urban emissions of the monitored cities.

Scattered radiance sensing is another usual approach in light pollution research. The artificial night sky brightness arises from the scattering of artificial light in the elementary volumes of the atmosphere located along the line of sight of the measuring device, from its input aperture to the top of the atmosphere, carrying information about the total light emissions. Several government, academic, and citizen science networks provide real-time measurements of the zenith sky radiance, using low-cost radiometers like the TESS-W (Zamorano et al. 2017, 2023; Pascual et al. 2021), SQM (Cinzano 2005; Bará 2016; Posch et al 2018), and others (Hänel et al. 2018; Alarcón et al. 2021). Unaided eye observations of the number of visible stars (Kyba et al. 2013, 2023) also belong to this category .

The radiance $L$ at the observer location (whether direct or scattered) is a linear function of the radiances $L_0$ of the sources (Falchi and Bará 2020). When $L$ and $L_0$ are strictly proportional, all other factors kept constant, their fractional changes turn out to be equal, $\Delta L/L = \Delta L_0/L_0$. It can then be expected that small, long-term, relative changes in the artificial emissions $L_0$ could be detected by analyzing large sets of measurements of $L$, after removing from them the remaining deterministic trends and by reducing as much as possible the uncertainties due to the variability of the atmosphere and the random noise of the detection system.

The challenge, not uncommon in environmental sciences, is to detect small inter-annual changes in the quantity of interest ($\Delta L_0/L_0$ of order $\sim$2%) by measuring the proxy variable $L$, which shows very large amplitude fluctuations in multiple time scales due to several deterministic processes and random perturbations which are not related to the long-term evolution of the emissions $L_0$ we wish to study.

The processes contributing to the deterministic variability of the radiance signal include, among others, the multi-scale cycles of natural light (daily, monthly, annual and inter-annual) from the Sun, Moon and other celestial sources (Kieffer and Stone 2005; Jones et al. 2013; Kocifaj 2009; Masana et al. 2021; Posch et al. 2018; Puschnig et al. 2014; Winkler 2022), changes in artificial light emissions throughout the night due to the varying human activity patterns (Dobler 2015; Bará et al. 2019), changes in the spectral composition of the sources (Sánchez de Miguel et al 2017; Bará et al 2019b), seasonal changes of the terrestrial albedo (Wallner and Kocifaj 2019), and detector sensitivity drifts due to aging (Puschnig et al. 2021;

Bará et al. 2021). Once the raw data have been filtered and these factors have been addressed and compensated for, the variability of the atmosphere plays a key role in limiting the minimum relative change $\Delta L_0/L_0$ that can be actually detected with a given confidence level by measuring $L$.

The atmosphere can be thought of as a variable medium interposed between the light source and the detector that perturbs the detector readings or, equivalently, as a variable component of the measurement system itself. The latter view is particularly appropriate for the scattered radiance approach, since the atmosphere provides the necessary constituents for enabling the scattered light paths connecting the source and the detector that are at the core of this method. The statistical variability of the molecular and aerosol optical depths (and other aerosol properties) decisively contribute to the combined uncertainty $u_L$ of the measurements, adding to the intrinsic detector noise.

Three approaches can be applied to reduce the uncertainty of the yearly estimates of $L$ caused by the variability of the atmosphere: (a) averaging large datasets of measurements made under a wide range of atmospheric conditions whose statistical distribution should not vary significantly from year to year, (b) making the measurements under approximately equal atmospheric conditions, or filtering them *a posteriori* in the raw datasets, (c) using reliable information on the instantaneous state of the atmosphere and appropriate theoretical models to retrieve $L_0$ from $L$ by correcting each individual measurement for its particular atmospheric state. While all these approaches can be applied alone or in combination to efficiently reduce the variability in $L$, they are not expected to succeed in zeroing the atmospheric contribution to the combined uncertainty. The reason in cases (b) and (c) is that the exact parameters of the atmosphere at the times of acquiring the measurements are seldom known in practice (although this difficulty can be circumvented to a large degree by using satellite atmospheric open data, see Puschnig et al 2023) and, in case (a), that the average statistical properties of the atmosphere may differ somewhat from year to year, due to multi-annual meteorological trends that may act as a strong confounding factor for detecting small emission changes.

The aim of this paper is to analyze the minimum changes in artificial light emissions that could be detected with a given confidence level by the direct and scattered radiance sensing approaches, as a function of the measurement geometry and of the residual variability of the molecular and aerosol optical depths. The results are expected to provide some useful hints for choosing the most convenient strategy to assess inter-annual changes of artificial light emissions for each particular user configuration.

## 2. Detecting small changes in artificial light emissions through a variable atmosphere

The possibility of detecting with a given confidence level a relative change of emissions $\Delta L_0/L_0$ by monitoring the change of the radiance at the observer location, $\Delta L/L = (L_2 - L_1)/L_1$, is limited by the uncertainty $u_{\Delta L}$ of $\Delta L$, the difference between the initial ($L_1$) and final ($L_2$) measured radiances.

For inter-annual monitoring, each of the terms $L_i$ ($i$=1,2) is calculated as the mean value of the radiances measured in the corresponding year. The set of measurements effectively used for this calculation may vary depending on the processing protocol chosen by the observers. They can use all available mesurements of that year or a subset of them, e.g. those corresponding to a given month or some pre-defined days, usually after filtering them for twilight, moonlight, clouds, and other confounding factors, in order to reduce the variability not due to changes in the actual emissions. Hencefort, the mentions in this paper to the statistical properties (mean values or uncertainties) of the radiance or of the state of the atmosphere in each year refer to those corresponding to the particular measurement set used for the calculation of $L_i$.

The uncertainty of $\Delta L$ is given by $u_{\Delta L} = (u_{L1}{}^2 + u_{L2}{}^2)^{1/2}$, where $u_{L1}$ and $u_{L2}$ are the uncertainties of the radiances $L_1$ and $L_2$, respectively. Each $u_{Li}$, in turn, is a combined uncertainty term, that in its basic form encompasses the standard deviation of the mean, $u_{s,Li}$ , and the intrinsic precision of the measuring device, $u_{p,Li}$ , such that $u_{Li} = \left(u_{s,Li}{}^2 + u_{p,Li}{}^2\right)^{1/2}$.

The $u_{\Delta L}$ described above is the relevant uncertainty to be used when it comes to detecting statistically significant changes $\Delta L$ in the radiances measured at the observer location. However, our final purpose is to detect changes $\Delta L_0$ in the actual emissions. To that end, the inter-annual variability of the atmosphere shall also be taken into account. If the average state of the atmosphere were exactly the same from year to year, then any significant difference $L_2 - L_1$ could be attributed to actual changes in the emissions, once the remaining confounding factors are removed or compensated for. However, the average state of the atmosphere does change from year to year, and this introduces an additional variability in $L_1$ and $L_2$ that is not related to emissions. Denoting by $u_{a,L}$ the standard deviation of the inter-annual set of mean values { $L_1$, $L_2$,...} due to inter-annual changes in the annual mean atmospheric parameters (as stated above this annual mean refers to the average atmospheric conditions in the set of measurements used to calculate $L_i$), the combined total uncertainty assigned to $L_i$ for the purpose of detecting changes in emissions is $u_{Li} = \left(u_{s,Li}{}^2 + u_{p,Li}{}^2 + u_{a,L}{}^2\right)^{1/2}$. Finally, if the measurements are made with the same or equivalent devices ($u_{p,L1} = u_{p,L2} \equiv u_{p,L}$) and the statistical variability of the mean is similar each year ($u_{s,L1} = u_{s,L2} \equiv u_{s,L}$), then $u_{L1} = u_{L2} \equiv u_L$, and hence $u_{\Delta L} = \sqrt{2}\, u_L$, with $u_L = \left(u_{s,L}{}^2 + u_{p,L}{}^2 + u_{a,L}{}^2\right)^{1/2}$.

The minimum change in $L$ that can be reliably detected is of order $\Delta L = K\, u_{\Delta L} = K \sqrt{2}\, u_L$, where $K$ is the coverage factor for calculating the expanded uncertainty associated to the desired confidence level. The precise value of $K$ depends on the underlying probability density function of the measurements, which is not always known. For many practical distributions, coverage factors in the range $K$=2 to $K$=3 provide confidence levels between ~95% and ~99%, respectively, when the number of deegrees of freedom is large. For the Gaussian probability density distribution, $K$(95.0%)=1.96, $K$(95.5%)=2, $K$(99.0%)=2.6 and $K$(99.7%)=3.

In this section we evaluate the uncertainty $u_{a,L}$ due to the residual variability from year to year of the mean molecular and aerosol optical depths of the atmosphere, and the fundamental limits that this imposes on the minimum detectable inter-annual $\Delta L$. Other terms contributing to $u_L$, like the detector precision $u_{p,L}$ and the statistical variability of the mean radiance within each year $u_{s,L}$, are strongly dependent on the details of each particular device and measurement protocol, and can be easily added afterwards. By restricting our study to the variability in $L$ due to atmospheric factors, $u_{a,L}$, the detection limits deduced in this paper shall be taken as an optimistic lower bound.

### *2.1 Basic overview*

It can be anticipated from basic principles that the variability of the state of the atmosphere should affect in a different way the measurement of the direct and scattered radiances. Let us consider, for the sake of the argument, an exceedingly simple situation: a homogeneous atmosphere of maximum height $H$, extinction coefficient $k$, and optical depth $\tau = k\,H$, with a single artificial light source located at F (Fig. 1). In such a medium, the direct radiance from the source reaching the observer O after propagating horizontally a distance $D$ scales with $k$ as $L_1 \sim \exp(-kD)$, due to the attenuation along the path FO. If the measurements of the direct radiance are made instead from an hypothetic satellite above the atmosphere, observing F with a nadir angle $n$, the radiance would scale with $k$ as $L_2 \sim \exp[-M(n)\,k\,H]$, where $M(n) \approx 1/\cos n$ is the airmass number.

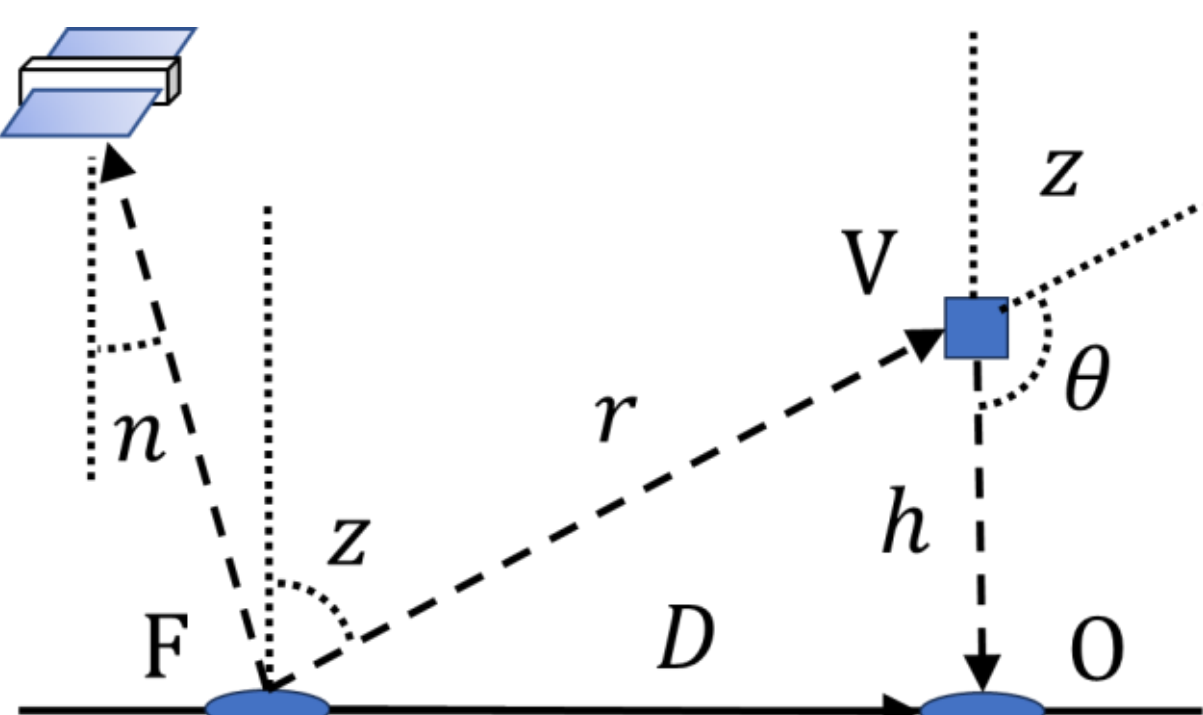

**Fig. 1.** Basic geometrical parameters for radiance sensing of night-time lights. F: artificial light source; O: ground-based observer; $D$: distance FO; V: elementary volume of the atmosphere at altitude $h$ in the vertical of O; $z$: zenith angle of V in the reference frame of F; $r$: distance FV; $\theta$: scattering angle at V of the light from F scattered towards the observer; $n$: nadir angle of the source in the satellite reference frame.

In turn, the zenith scattered radiance reaching O due to the light that underwent a single scattering event in the elementary atmospheric volume V scales with $k$ as $L_3 \sim k\exp(-kD')$, where $D' = r + h$. This expression results from the combination of the amount of scattered light at V, proportional to $k$, and the attenuation along the path FVO, proportional to the exponential term. Note that the fluctuations in $k$ affect in opposite ways the scattering and the attenuation, which are complementary phenomena with competing effects. For different observation geometries and values of the intervening parameters, an increase in $k$ (e.g. by an increase in the aerosol content of the atmosphere) could result in an increase or a reduction of the artificial zenith radiance seen by O, depending on whether the scattering or the attenuation prevails, or even in no change at all, if the change in one factor exactly compensates the opposite change in the other. It must be kept in mind that the path FVO is only one among the multiple paths contributing to the build up of the zenith radiance at O: the total radiance is obtained by adding up the contributions of the light travelling from the source to every elementary volume in the vertical of O (from $h = 0$ to $h = H$), undergoing there a single scattering event, and being re-directed towards the ground observer. Each of these paths has its own value of $D'$.

Denoting by $\sigma_k$ the standard deviation of the fluctuations of $k$, assumed to be small, the uncertainties in the determination of the radiances scale with $k$ as $u_1 \sim D \exp(-kD)\, \sigma_k$, $u_2 \sim M(n)\, H \exp[-M(n)\, k\, H]\, \sigma_k$, and $u_3 \sim |1 - kD'| \exp(-kD')\, \sigma_k$, respectively. The relative uncertainties are then $u_1/L_1 \sim D\, \sigma_k$, $u_2/L_2 \sim M(n)\, H\sigma_k$, and $u_3/L_3 \sim |(1/k) - D'|\, \sigma_k$. Thus, while the relative uncertainty of the measurements made along the direct horizontal path is expected to increase linearly with the distance from the observer to the source, for the satellite observations it increases with the observation nadir angle via the airmass number, and for the radiance scattered at V observed by O it depends, among other factors, on by how much $1/k$ differs from $D'$. Recall that $1/k$ is the mean free path of photons in this homogeneous atmosphere. When these two terms are equal, $1/k = D'$, the contribution of the small fluctuations of $k$ to the uncertainty $u_3$ is zero, because any first-order change in the amount of scattering at V is exactly compensated by an opposite change in the overall transmittance of the path FVO.

It can then be expected that some of these radiance sensing approaches may be more advantageous than others for detecting small changes in artificial light emissions, depending on the detection configuration, the variability of the atmosphere, and the distance between the source and the observer.

The above reasoning is overly simplified, and it is only intended for introducing some of the elementary physical processes at play. A more detailed calculation, based on well-known single-scattering models with artificial light sources is outlined in subsection 2.2.

### *2.2 Minimum detectable changes in artificial light emissions*

Several single- and multiple-scattering models with ground-level artificial light sources have been developed in the last decades to study the propagation of light pollution in the terrestrial atmosphere (e.g. Garstang 1989; Cinzano and Falchi 2012; Kocifaj 2007, 2018; Aubé and Simoneau 2018; Simoneau et al. 2021). The suitability of using a single-scattering model is generally restricted to short (a few km) distances in not very thick atmospheres, since otherwise higher scattering orders become relevant and may even be dominant (Kocifaj et al. 2024). For the purposes of this paper a single-scattering model is expected to provide useful results, since the evaluation of the long-term trends in artificial light emissions is generally attempted using measurements collected in clear nights from sites not far from the main cities, or by using radiometers onboard satellites measuring the light sources through not very large airmasses.

To that end, let us consider an atmosphere composed of molecular (R) and aerosol (A) constituents, whose wavelength-dependent extinction coefficients $k_R$ and $k_A$ decrease exponentially with the altitude $h$ with scale heights $H_{\mathrm{R}}$ and $H_{\mathrm{A}}$, respectively, such that

$$k(h,\lambda) = \sum_{i=\mathrm{R,A}} k_i(h,\lambda) = \sum_{i=\mathrm{R,A}} k_{0i}(\lambda) \exp(-h/H_i) \tag{1}$$

The associated optical depths are $\tau_i(\lambda) = \int_0^{\infty} k_i(h,\lambda)\, \mathrm{d}h = k_{0i}(\lambda)\, H_i,\ (i = \mathrm{R,A})$. The spectral dependence of the molecular and aerosol optical depths is generally described by different approximate power laws ($\sim 1/\lambda^{\alpha}$), with molecules showing a typical Rayleigh behavior ($\alpha \approx 4$) and aerosols with smaller Angström exponents ($\alpha \approx 1$). For the purposes of this work it is advantageous to express the optical depth at wavelength $\lambda$ as a function of its value at an arbitrary reference wavelength $\lambda_0$, such that $\tau_i(\lambda) = \tau_{0i}\, f_i(\lambda)$, where $\tau_{0i} \equiv \tau_i(\lambda_0)$ and the $f_i(\lambda), (i = \mathrm{R,A})$, are the functions describing the relative wavelength dependence (for particular choices of these funtions see section 3).

Without loss of generality for the calculations of this section, it can assumed that the spectral source radiance has azimuthal symmetry, $L_0(z,\lambda)$, being only dependent on the zenith angle, $z$, but not on the azimuth. The spectral sensitivity of the detection system (detector passband) will be denoted by $S(\lambda)$.

Under these assumptions, using the notation of Fig. 1, the direct radiance $L_{\mathrm{dir}}$ detected by the observer after propagating a distance $D$ along the horizontal path FO is:

$$L_{\mathrm{dir}} = \int_{\Lambda} S(\lambda)\, L_0\left(\frac{\pi}{2},\lambda\right) \exp\left(-\sum_{i=\mathrm{R,A}} \frac{\tau_{0i}\, f_i(\lambda)}{H_i} D\right) d\lambda \tag{2}$$

where $\Lambda$ is the relevant range of wavelengths over which the spectral integration shall be carried out, formally and by default $\Lambda = (0,\infty)$. The uncertainty in $L_{\mathrm{dir}}$ due to the fluctuations of $\tau_{0\mathrm{R}}$ and $\tau_{0\mathrm{A}}$, with standard deviations $\sigma_{\tau_{0\mathrm{R}}}$ and $\sigma_{\tau_{0\mathrm{A}}}$, respectively, is $u_{L_{\mathrm{dir}}} = \left[\sum_{j=\mathrm{R,A}} \left(\partial L_{\mathrm{dir}}/\partial \tau_{0j}\right)^2 {\sigma_{\tau_{0j}}}^2\right]^{1/2}$, with explicit expression:

$$u_{L_{\mathrm{dir}}} = D\left\{\sum_{j=\mathrm{R,A}} \left[\int_{\Lambda} S(\lambda)\, L_0\left(\frac{\pi}{2},\lambda\right) f_j(\lambda) \exp\left(-\sum_{i=R,A} \frac{\tau_{0i}\, f_i(\lambda)}{H_i} D\right) \mathrm{d}\lambda\right]^2 \frac{{\sigma_{\tau_{0j}}}^2}{{H_j}^2}\right\}^{1/2} \tag{3}$$

Evaluating the minimum detectable change in $L_{\mathrm{dir}}$ by using an expanded uncertainty with coverage factor $K$, such that $\Delta L_{\mathrm{dir}} = K\sqrt{2}\, u_{L_{dir}}$, the minimum relative change detectable with this approach is:

$$\frac{\Delta L_{\mathrm{dir}}}{L_{\mathrm{dir}}} \geq K\sqrt{2}\, \frac{u_{L_{\mathrm{dir}}}}{L_{\mathrm{dir}}} \tag{4}$$

where the values of $L_{\mathrm{dir}}$ and $u_{L_{\mathrm{dir}}}$ are numerically evaluated from Eqs. (2) and (3), respectively. The basic physical behavior underlying these equations can be revealed by analytically evaluating them for the limiting case of quasi-monochromatic or narrow-band detection at wavelength $\lambda_0$, $S(\lambda) = \delta(\lambda - \lambda_0)$, where the symbol $\delta$ stands for Dirac-delta distribution, obtaining:

$$\left(\frac{\Delta L_{\mathrm{dir}}}{L_{\mathrm{dir}}}\right)_{\lambda_0} \geq K\sqrt{2}\, D \sqrt{\frac{{\sigma_{\tau_{0\mathrm{R}}}}^2}{{H_{\mathrm{R}}}^2} + \frac{{\sigma_{\tau_{0\mathrm{A}}}}^2}{{H_{\mathrm{A}}}^2}} \tag{5}$$

which suggests that the ability for detecting small radiance changes will worsen if the distance $D$ between the detector and the sources is increased, and/or if the residual variability of the molecular and aerosol optical depths weighted by the inverses of their respective scale heights is larger.

Artificial light emissions are also monitored by radiometers onboard low Earth orbit satellites. The expressions for the direct radiance $L_{\mathrm{sat}}$ detected by a satellite along a slanted path at a nadir observation angle $n$, and its associated uncertainty $u_{\mathrm{sat}}$ due to atmospheric fluctuations, are:

$$L_{\mathrm{sat}} = \int_{\Lambda} S(\lambda)\, L_0(n,\lambda) \exp\left(-M(n) \sum_{i=\mathrm{R,A}} \tau_{0i}\, f_i(\lambda)\right) \mathrm{d}\lambda \tag{6}$$

$$u_{\mathrm{sat}} = M(n) \left\{ \sum_{j=\mathrm{R,A}} \left[ \int_{\Lambda} S(\lambda)\, L_0(n,\lambda)\, f_j(\lambda) \exp\left(-M(n) \sum_{i=\mathrm{R,A}} \tau_{0i}\, f_i(\lambda)\right) \mathrm{d}\lambda \right]^2 {\sigma_{\tau_{0j}}}^2 \right\}^{1/2} \tag{7}$$

where $M(n) \approx 1/\cos n$ is the airmass number (for more accurate expressions of the airmasses see section 3). The emissions change detection limit is given by an expression similar to Eq.(4) using the appropriate quantities, Eqs. (6)-(7). Again, some basic insights about the behavior of these equations can be obtained by evaluating the limiting case of quasi-monochromatic or narrow-band detection at wavelength $\lambda_0$, which gives:

$$\left(\frac{\Delta L_{\mathrm{sat}}}{L_{\mathrm{sat}}}\right)_{\lambda_0} \geq K\, \sqrt{2}\, M(n) \sqrt{{\sigma_{\tau_{0\mathrm{R}}}}^2 + {\sigma_{\tau_{0\mathrm{A}}}}^2} \tag{8}$$

Equation (8) suggests that the most sensitive detection of light emission changes can be achieved in near-nadir view, $M(n \approx 0) \approx 1$, being limited by the atmospheric variability.

The equations for ground-based radiometers monitoring the artificial night sky brightness are a bit more complex. The single-scattering zenith radiance $L_{\mathrm{sca}}$ at O due to the emissions of an elementary horizontal area $\mathrm{d}F$ of the source F is, following the general approach of Kocifaj (2007):

$$L_{\mathrm{sca}} = \mathrm{d}F \int_{\Lambda} S(\lambda)\, L_0(z,\lambda) \int_{h=0}^{\infty} \frac{\cos z}{r^2} \left[ \sum_{i=\mathrm{R,A}} w_i(\lambda) \frac{\tau_{0i}\, f_i(\lambda)}{H_i} \exp\left(\frac{-h}{H_i}\right) p_i(\theta,\lambda) \right]$$
$$\times \exp\left\{ - \sum_{i=\mathrm{R,A}} [M(z)+1]\, \tau_{0i}\, f_i(\lambda) \left[1-\exp\left(\frac{-h}{H_i}\right)\right] \right\} \mathrm{d}h\ \mathrm{d}\lambda \tag{9}$$

where $r = (h^2 + D^2)^{1/2}$, $\cos z = h/r$, $\theta = \pi - z$ is the scattering angle for the geometry of Fig. 1, $M(z) = 1/\cos z$, and $w_i(\lambda)$ are the single scattering albedos (SSA) of molecules and aerosols (for molecules, $w_{\mathrm{R}} = 1$). The $p_i(\theta,\lambda)$ are the scattering phase functions accounting for the angular distribution of the light scattered by molecules and aerosols, normalized to 1 when integrated over the whole $4\pi$ sr solid angle space of scattering angles $\theta$:

$$p_{\mathrm{R}}(\theta,\lambda) \equiv p_{\mathrm{R}}(\theta) = \frac{1}{4\pi} \times \frac{3}{4}(1+\cos^2\theta) \tag{10}$$

$$p_{\mathrm{A}}(\theta,\lambda) = \frac{1}{4\pi} \times \frac{1-g(\lambda)^2}{[1+g(\lambda)^2 - 2g(\lambda)\cos\theta]^{3/2}} \tag{11}$$

where $p_{\mathrm{R}}(\theta)$ has the classic Rayleigh angular dependence, and a generic Henyey-Greenstein aerosol phase function has been chosen for $p_{\mathrm{A}}(\theta,\lambda)$. The uncertainty due to the variability of the atmospheric optical depths is then given by $u_{L_{\mathrm{sca}}} = \left[\sum_{j=\mathrm{R,A}} \left(\partial L_{\mathrm{sca}}/\partial \tau_{0j}\right)^2 \sigma_{\tau_{0j}}{}^2\right]^{1/2}$, where

$$\frac{\partial L_{\mathrm{sca}}}{\partial \tau_{0j}} = \mathrm{d}F \int_{\Lambda} S(\lambda)\, L_0(z,\lambda) \int_{h=0}^{\infty} \frac{\cos z}{r^2} \Bigg( \left[ w_j(\lambda) \frac{f_j(\lambda)}{H_j} \exp\left(\frac{-h}{H_j}\right) p_j(\theta,\lambda) \right]$$
$$- \left[ \sum_{i=\mathrm{R,A}} w_i(\lambda) \frac{\tau_{0i}\, f_i(\lambda)}{H_i} \exp\left(\frac{-h}{H_i}\right) p_i(\theta,\lambda) \right] [M(z)+1]\, f_j(\lambda) \left[1-\exp\left(\frac{-h}{H_j}\right)\right] \Bigg)$$
$$\times \exp\left\{ - \sum_{i=\mathrm{R,A}} [M(z)+1]\, \tau_{0i}\, f_i(\lambda) \left[1-\exp\left(\frac{-h}{H_i}\right)\right] \right\} \mathrm{d}h\ \mathrm{d}\lambda \tag{12}$$

and, finally, the minimum detectable relative change in the zenith scattered radiance is

$$\frac{\Delta L_{\mathrm{sca}}}{L_{\mathrm{sca}}} \geq K\sqrt{2}\, \frac{u_{L_{\mathrm{sca}}}}{L_{\mathrm{sca}}} \tag{13}$$

which shall be numerically evaluated using Eqs. (9) and (11), since it has no simple analytical expression even for the limiting case of quasi-monochromatic or narrowband detection, due to the remaining internal integration over $h$ along the whole air column.

Let us recall that for monitoring inter-annual changes in emissions, the measured radiances $L$ in the above expressions should be interpreted as annual average values, whose year-to-year evolution $\Delta L$ is to be determined for evaluating from $\Delta L/L$ the changes $\Delta L_0/L_0$. The combined standard deviation of $L$, under the assumption of no changes in actual emissions, encompasses two contributing terms. One of them is the standard deviation assigned to *each* individual yearly mean, based on the measurements made that year and the precision of the measuring device.

This term can be substantially reduced by averaging a large number of measurements, enabling a very effective averaging of the seasonal atmospheric conditions and resulting in a very low type A uncertainty for that annual mean (whose total uncertainty would then be mostly determined by type B factors like the detector precision). The other term contributing to the uncertainty of $L$ is the standard deviation of the inter-annual set of annual means, not due to actual emission changes. The mean values of $L$ in different years are not exactly equal to each other –even for constant emissions– due to the unavoidable inter-annual fluctuations in the average atmospheric conditions. In particular, the mean values of $\tau_{0\mathrm{R}}$ and $\tau_{0\mathrm{A}}$ fluctuate from year to year due to periodic meteorological oscillations, variable biomass burning and dust emissions, and long-term climate trends, among other factors. Even if the measurements made each year are filtered to select those acquired under approximately uniform atmospheric conditions, some degree of residual fluctuation in the filtered mean values of $\tau_{0\mathrm{R}}$ and $\tau_{0\mathrm{A}}$ for different years will remain. Whatever the case, the corresponding inter-annual standard deviations $\sigma_{\tau_{0\mathrm{R}}}$ and $\sigma_{\tau_{0\mathrm{A}}}$ will limit, via the equations in Section 2, the minimum detectable emission changes. They induce a variability in $\Delta L$ which is not due to intrinsic variations of the emissions, $\Delta L_0$, but to confounding atmospheric effects.

## 3. Results

The smallest light emission changes that may be detected, according to Section 2, depend on multiple geometric and atmospheric parameters. To get some insights about their order of magnitude we show in the present section the behavior of $\Delta L/L$ versus $D$ for some particular choices. We consider an atmosphere with scale-heights $H_{\mathrm{R}} = 8$ km and $H_{\mathrm{A}} = 1.5$ km, molecular (Rayleigh) optical depth $\tau_{0\mathrm{R}}(@550\ \mathrm{nm}) = 0.10$, and several values of the aerosol optical depth $\tau_{0\mathrm{A}}(@550\ \mathrm{nm}) = [0.05, 0.1, 0.2, 0.3]$. The wavelength dependence of the molecular optical depth for the standard atmosphere was taken as $\tau_{\mathrm{R}}(\lambda) = 0.00879\ \times \lambda^{-4.09}$, with $\lambda$ expressed for this formula in micrometers (Teillet, 1990), so that $f_{\mathrm{R}}(\lambda) = (\lambda/\lambda_0)^{-4.09}$ with $\lambda$ and $\lambda_0$ in consistent units . The wavelength dependence of the aerosol optical depth, $\tau_{\mathrm{A}}(\lambda)$, was calculated from the reference value $\tau_{0\mathrm{A}}(@550\ \mathrm{nm})$ by means of equation (6) of McComiskey et al (2008), from which $f_{\mathrm{A}}(\lambda)$ can be immediately identified. The single-scattering albedos (SSA) were set to $w_{\mathrm{R}} = 1$ and $w_{\mathrm{A}}(@550\ \mathrm{nm}) = 0.85$. The aerosol assymmetry parameter (ASY) at the reference wavelength was chosen as $g(@550\ \mathrm{nm}) = 0.7$. The wavelength dependence of the aerosol SSA and ASY was accounted for by using equations (7)-(8) of McComiskey et al (2008), with Angström scattering and absorption exponents both equal to 1.0. The airmasses for satellite nadir angles $\beta$ were calculated as $M(\beta) = 1/(\cos\beta + 0.50572\ \times\ (96.07995° - \beta)^{-1.6364})$ according to Kasten and Young (1989). The geometrical factor for scattered light calculations is $M(z) = 1/\cos z$.

Two detector spectral bands $S(\lambda)$ were chosen for this calculation, depending on the measurement approach. The measurements of the direct and scattered radiances by the ground-based observer were assumed to be made by means of the TESS-W radiometer (Zamorano et al. 2017; Pascual et al. 2021) whose spectral passband $S(\lambda)$ is described in Bará et al. (2019). The direct radiances measured from low Earth orbit were evaluated, in turn, for the spectral passband of the widely used VIIRS-DNB radiometer onboard the satellite Suomi-NPP (Elvidge et al. 2017, 2022; Román et al. 2018). The spectral integrations were carried out for the wavelength range 350-1000 nm, with discrete 5 nm steps.

The outdoor lamps mix was chosen as composed by 80% high pressure sodium (CCT≈2200 K) and 20% LED (CCT=4000 K), where the percents refer to the contribution of each type of lamp to the total emitted lumen (lm). Their individual spectra are displayed in figure 3 of Bará et al (2019b). The combined lamp spectrum was assigned to $L_0(z,\lambda)$ for the direct radiance measurements made by the ground-based observer in Eqs. (2)-(3). Satellite radiometers, especially at small nadir angles, seldom record the direct light from the lamps, but rather the radiance reflected by the ground. So, the spectral radiance $L_0(n,\lambda)$ for the satellite measurements, Eqs. (6)-(7), was taken as proportional to the product of the lamp radiance and the bi-directional reflectance distribution function (BRDF) of the terrain, here assumed to be Lambertian and with the spectral reflectance of a mixture of asphalt and two types of concrete samples from the USGS Spectral Library (Kokaly et al. 2017) in proportions 1:0.25:0.25. Their individual reflectances are displayed in figure 4 of Bará et al. (2019b). Finally, for the zenith scattered radiance measurements in Eqs. (9) and (12), it was assumed that the artificial light sources emitted 10% of total lumen towards the upper hemisphere (with the spectrum of the lamp mix) whereas the remaining 90% were emitted towards the ground and reflected with the same BRDF as for the satellite measurements, being $L_0(z,\lambda)$ the sum of these two terms.

As commented in section 2, the standard deviations $\sigma_{\tau_{0\mathrm{R}}}$ and $\sigma_{\tau_{0\mathrm{A}}}$ refer here to the variability of the annual mean values of $\tau_{0\mathrm{R}}$ and $\tau_{0\mathrm{A}}$ across the sample of years for which the study is made, regardless of whether this variability is due to actual meteorological trends or to residual fluctuations present in the annual averages after selecting the measurements made each year under approximately constant atmospheric conditions. The actual values of $\sigma_{\tau_{0\mathrm{R}}}$ and $\sigma_{\tau_{0\mathrm{A}}}$ are strongly dependent on the particular set of years included in the study, the measurement procedures and the filtering algorithms, and it is expected that they can be very different across all possible applications. Some particular choices have been made here based on the data below.

The molecular (Rayleigh) optical depth $\tau_{\mathrm{R}}$ is proportional to the atmospheric pressure (Bodhaine et al. 1999; Dutton en al. 1994; Teillet 1990; Young 1981). The standard pressure at sea level is 1013.25 hPa, with fluctuations that seldom reach $\pm 50$ hPa above or below this value. The inter-annual standard deviation of the annual mean sea level pressure depends on several factors, including the phase of the atmospheric oscillation cycles, the latitude of the observers, and other features (van Loon and Madden 1983; Thompson and Wallace 2001; Wallace and Thompson 2002). For reference we have taken a typical inter-annual standard deviation $\sigma_P = 5$ hPa from van Loon and Madden (1983). This corresponds to a relative pressure variability $\sigma_P/P \approx 0.005$. Since $\sigma_{\tau_{0\mathrm{R}}}/\tau_{0\mathrm{R}} \approx \sigma_P/P$, this translates into $\sigma_{\tau_{0\mathrm{R}}} \approx 5 \times 10^{-4}$ for the assumed $\tau_{0\mathrm{R}}(@550\ \mathrm{nm}) = 0.10$. For the calculations in this section we have also included two larger additional values, such that $\sigma_{\tau_{0\mathrm{R}}}(@550\ \mathrm{nm}) = [0.0005, 0.001, 0.002]$.

The aerosol optical depth at any location typically shows large fluctuations across the year, and also within any given day or night (Benavent-Oltra et al. 2019; Cavazzani et al. 2020; Fernández et al. 2017; Hess et al. 1998; Horvath et al. 2002; Puschnig et al. 2021, 2023; Román et al. 2017). Daily data on aerosols at multiple observing sites are available from AERONET (Giles et al. 2019). As an example of the intervening orders of magnitude, for the city of A Coruña (Galicia, Spain, European Union), the AERONET records from 2012 until 2015 in periods without dust episodies show that the mean aerosol optical depth at wavelength $\lambda$ =442 nm was $\tau_{\mathrm{A}}$= 0.27$\pm$0.08, and at $\lambda$ =667 nm it was $\tau_{\mathrm{A}}$= 0.14$\pm$0.05, where the standard deviations of the data were derived from the InterQuartile Range 25%–75% (digitized from figure 11 of Fernández et al. 2017) assuming an underlying Gaussian distribution. A recent review of the uncertainties in aerosol optical depths estimated with different methods can be found in Vogel et al. (2022). Analyzing the aerosol optical depth datasets acquired by an ensemble of satellites between 60°N and 60°S in the period 1998–2019, at mid-visible wavelengths, these authors reported a spatial mean $\tau_{\mathrm{A}} = 0.140 \pm 0.018$ (standard deviation of the ensemble), with an average year-to-year variability of 0.00394 (arithmetic mean of the year-to-year standard deviations from the individual satellites). For the calculations in the present section we have adopted that value and two larger ones, $\sigma_{\tau_{0\mathrm{A}}}(@550\ \mathrm{nm}) = [0.004, 0.01, 0.02]$.

The results are displayed in Figure 2. The Y-axes show the minimum detectable $\Delta L/L$ for a confidence level ~95% (expanded uncertainty with $K = 2$), and the X-axes the horizontal distances $D$ from the source F to the observer O, in km. For satellite observations, $D$ corresponds to the distance between the source and the satellite nadir ground point. Each individual panel in Fig. 2 corresponds to a different combination of $\sigma_{\tau_{0\mathrm{R}}}$ and $\sigma_{\tau_{0\mathrm{A}}}$, with $\sigma_{\tau_{0\mathrm{R}}}(@550\ \mathrm{nm}) = [0.0005, 0.001, 0.002]$ increasing from top to bottom, and $\sigma_{\tau_{0\mathrm{A}}}(@550\ \mathrm{nm}) = [0.004, 0.01, 0.02]$ increasing from left to right. The different sets of curves in each panel correspond to the three approaches analyzed in this work: direct radiance measured by the ground observer, $L_{\mathrm{dir}}$, direct radiance measured by the satellite, $L_{\mathrm{sat}}$, and zenith scattered radiance measured from ground, $L_{\mathrm{sca}}$. For each approach four individual curves corresponding to a single molecular optical depth $\tau_{0\mathrm{R}}(@550\ \mathrm{nm}) = 0.10$ and four aerosol optical depths $\tau_{0\mathrm{A}}(@550\ \mathrm{nm}) = [0.05, 0.1, 0.2, 0.3]$ are drawn. For $L_{\mathrm{dir}}$, and also for $L_{\mathrm{sat}}$, the four curves are practically superimposed onto each other and are hardly distinguishable, which means that for these methods the relative emission changes that can be detected do not depend critically on the average state of the atmosphere, only on its variability. The curves for $L_{\mathrm{sca}}$ show some spread of values, especially for short and mid-distances, depending on $\tau_{0\mathrm{R}}$ and $\tau_{0\mathrm{A}}$.

Several trends can be observed in these plots. As anticipated in section 2.1, the minimum detectable change $\Delta L/L$ by direct $L_{\mathrm{dir}}$ radiance sensing increases almost linearly with the distance $D$, suggesting that this method may provide the best results at short distances from the sources. The performance of the satellite measurements, in turn, remains practically constant for the range of distances here considered. As a matter of fact the $\Delta L/L$ for $L_{\mathrm{sat}}$ actually increases with the nadir angle (which increases with $D$) but this effect is hardly visible in these panels, because the Suomi-NPP satellite orbits Earth at a nominal altitude of 824 km, the nadir observation angle for the range of distances displayed in Fig. 2 never surpasses the value $n = \tan^{-1}(100/824) \approx 7°$, and hence the airmass number $M(n)$ lies in the narrow range 1.0000–1.0075. It is worth recalling that many VIIRS-DNB data are acquired at larger nadir angles, for which $M(n)$ is noticeably greater than 1, so the situation represented in Fig. 2 only applies to near-nadir measurements. The curves of $\Delta L/L$ for $L_{\mathrm{sca}}$ show minima at some distances from the sources. Each of these $L_{\mathrm{sca}}$ curves stems from the sum of multiple individual contributions from the elementary d$h$ and d$\lambda$ intervals of the integrand of Eq. (12), and each of these contributions is expected to have a minimum located at a given value of $D$ (see section 2.1). For some configurations of the residual variability ($\sigma_{\tau_{0R}}, \sigma_{\tau_{0A}}$) this method allows to detect the smallest $\Delta L/L$, outperforming the satellite measurements (for a detailed analysis of this issue, based on observational data, see Wallner et al, 2023).

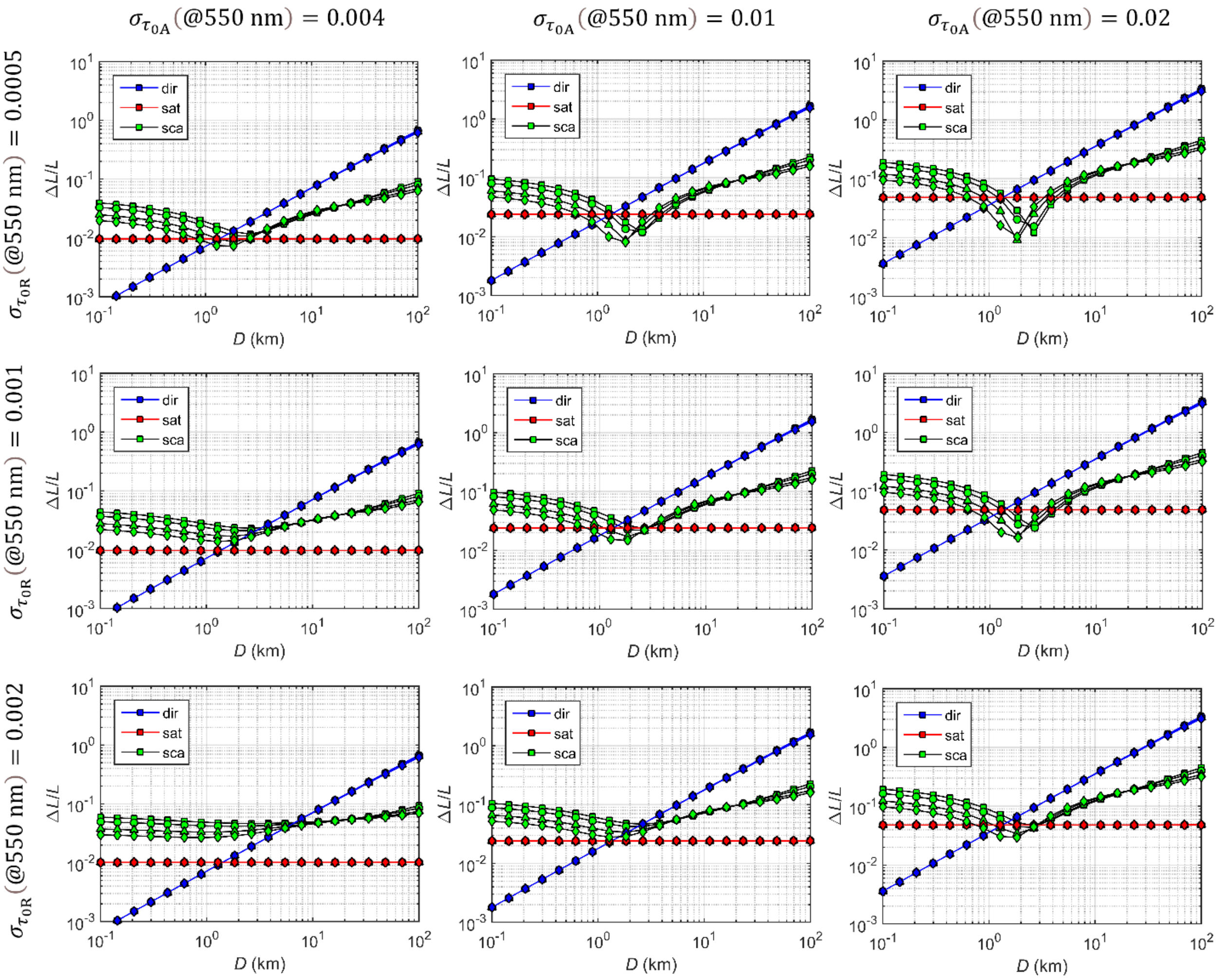

**Fig. 2.** Minimum relative changes $\Delta L/L$ detectable with a confidence level ~95% ($K = 2$) for the direct radiance measured by the ground observer, $L_{\mathrm{dir}}$, the direct radiance measured from the satellite, $L_{\mathrm{sat}}$, and the zenith scattered radiance measured from gound, $L_{\mathrm{sca}}$, versus the horizontal distance $D$ from the source F to the observer O, in km. For satellite measurements $D$ is the distance between the source and the satellite nadir ground point. Each individual panel corresponds to a different combination of $\sigma_{\tau_{0R}}$ and $\sigma_{\tau_{0A}}$, with $\sigma_{\tau_{0R}}(@550\ \mathrm{nm}) = [0.0005, 0.001, 0.002]$ increasing from top to bottom, and $\sigma_{\tau_{0A}}(@550\ \mathrm{nm}) = [0.004, 0.01, 0.02]$ increasing from left to right. Four curves are plotted for each measurement approach, corresponding to different mean aerosol optical depths at 550 nm: 0.05

(squares), 0.1 (circles), 0.2 (triangles) and 0.3 (diamonds). The curves are practically superimposed for the direct and satellite measurement approaches. Additional details in the text.

The possibility of detecting small changes $\Delta L/L$ with all measurement strategies worsens, as expected, for larger values of the residual variability of the atmosphere. This is particularly noticeable by comparing the columns of Fig. 2 from left to right, since the values of $\sigma_{\tau_{0\mathrm{A}}}$ dominate the global variability in our example. The values of $\sigma_{\tau_{0\mathrm{R}}}$ (which increase from top to bottom) are about one order of magnitude smaller. It can be seen that the minimum $\Delta L/L$ for $L_{\mathrm{dir}}$ and $L_{\mathrm{sat}}$ practically does not change from one row to the next. However, the effect of increasing $\sigma_{\tau_{0\mathrm{R}}}$ on $L_{\mathrm{sca}}$ can be seen in the progressively higher values of the minima of its curves for each constant value of $\sigma_{\tau_{0\mathrm{A}}}$.

## 4. Discussion

To put in context these results, let us recall that fractional changes $\Delta L_0/L_0$ in the source emissions give rise to equal or very similar changes $\Delta L/L$ in the radiances reaching the observer. Figure 2 displays the smallest $\Delta L/L$ that could be detected at a confidence level 95%, due to the uncertainty caused by the atmospheric variability of the molecular and aerosol optical depths.

The $L_{\mathrm{dir}}$ sensing approach provides better results than the other two options for very short distances to the sources (~0.9%–3% at 1 km), but its performance worsens in proportion to the distance, quickly reaching levels above 10%. This can be an appropriate method for monitoring the emissions of small isolated towns (smaller than ~1 km wide), if there are vantage points from which the city lights (or their diffuse reflections on façades and pavements) can be directly imaged from sufficiently short $D$. For larger population nuclei, however, many light sources will be located at distances that preclude their precise monitoring by this approach.

Satellite ($L_{\mathrm{sat}}$) sensing provides in general the most stable and precise determination of emission changes for distances larger than ~1km, with limits of order 1% for small residual $\sigma_{\tau_{0A}}$ and 4% for larger aerosol variabilities. Note however that this stable precision is due to the small nadir angles involved, which closely correspond to a 1.0 airmass across the whole range of $D$, and that it will deteriorate as the airmass number increases. Satellite radiance sensing allows monitoring large areas of the world on a daily basis, outperforming in this aspect its counterparts, which are limited to local or regional coverage. This advantange comes however at a price, namely, satellite radiometers gather considerably less measurement samples per year (for any given site on Earth) than ground-based stations measuring $L_{\mathrm{dir}}$ or $L_{\mathrm{sca}}$, which can continuously record local radiances at very fast rates.

Finally, $L_{\mathrm{sca}}$ shows an interesting behavior: within a specific range of distances it provides better detection capabilities than $L_{\mathrm{dir}}$ or $L_{\mathrm{sat}}$, with a performance that mostly depends on $\sigma_{\tau_{0R}}$, being relatively insensitive to $\sigma_{\tau_{0A}}$. However, the $\Delta L/L$ limit quickly increases outside this range. This may preclude using this method for monitoring large metropolitan areas, where a significant fraction of the sources may lie outside the optimum distance interval.

The inter-annual variability of the molecular and aerosol optical depths is only one of the many factors determining the measurement variability not due to actual changes in the source emissions. Inter-annual fluctuations in the mean value of other atmospheric parameters like the aerosol albedo, Ängstrom exponents for absorption and scattering, or the asymmetry factor (Dubovik et al. 2002; Horvath et al. 2002; Korras-Carraca et al. 2015) would add to it. Other uncertainty sources as e.g. the sensor precision or the inter-annual fluctuations of the average nadir angle of satellite measurements will further limit the ability to detect small emission trends. The results shown in Section 3 should then be understood as lower bounds. Overall, this highlights the need of using efficient data filtering and processing algorithms, and of developing low-noise detectors with proven long-term (multi-year) stability, in order to reduce as much as possible the uncertainty of the annual radiance averages. This is of particular importance if overall precisions of order 1% are intended to be achieved for monitoring the evolution of antropogenic light emissions on a yearly basis.

The detected changes in the emitted radiance $L_0$ (units W $m^{-2}$ $sr^{-1}$) can be used to calculate the changes in the average radiant exitance (W $m^{-2}$) of the monitored area if the angular emission function of the sources is known. The exitance is the spatial density of emitted radiant flux $\Phi_0$ (W). Since the radiance is an energy density per unit time, solid angle, and area perpendicular to the direction of propagation, it is a direction-dependent physical quantity. The direct radiance $L_{\text{dir}}$ only provides information (after accounting for propagation losses) about the radiance $L_0(z, \varphi; \lambda)$ emitted by the source in the specific direction $(z_0, \varphi_0)$ joining the source and the observer. In order to calculate $\Phi_0$ from $L_0$ it is necessary to add up angularly the contributions of the radiance emitted in all directions $(z, \varphi)$, weighted by $\cos z$. This requires knowing the shape of the angular emission function of the source, either by means of field measurements (Kocifaj, 2017), educated guesses based on first principles (Kocifaj, 2018b), or numerical calculations based on the city structure (Espey, 2021). The specific shape of the angular emission function does not affect, however, to the assessment of relative changes in the overall radiant flux from one year to the next, as long as this shape is kept constant over time. Note that, unlike $L_{\text{dir}}$, each annual dataset of $L_{\text{sat}}$ contains information about the radiance emitted by the source in a discrete set directions (determined by the satellite observation angles), and every single measurement of $L_{\text{sca}}$ contains aggregated information on the radiance emitted by the source in the continuous range of directions that intersect the observer's line of sight. The findings of this paper would not necessarily contradict the (average) estimate of the 2.2% emissions increase per year based on satellite measurements, but could substantially modify the estimation of its uncertainty, and hence of its significance. This is a topic that deserves a careful analysis by combining the satellite radiance measurements with a simultaneous set of atmospheric data, in particular about the molecular and aerosol optical depths.

Complementary to the described radiance measurement approaches, administrative records may provide relevant information. Inventories of outdoor lights are increasingly available in many municipalities. They are usually elaborated as part of public services surveys, and published to comply with the tenets of trasparency and freedom of information of local administrations. They may also be a required document when applying for public funds for streetlighting remodeling. Some of these inventories provide comprehensive information even at the micro-scale, with multiple data fields for each individual light source installed on the streets. An open access example is the Paris public lighting website (Paris, France, 2024). In general, this kind of inventories do not include all outdoor light sources, but only those belonging to the public streetlighting system, which in many countries comprise a large fraction of the total. Other public or publicly authorized lights (e.g. commercial and industrial lighting, LED billboards, and other regulated displays) are often recorded in separate databases, but in principle nothing prevents combining them in a single registry. Two obvious caveats regarding administrative lighting records are whether they are updated with sufficient frequency and whether their accuracy is double-ckecked by independent control bodies. The amount of emissions of unregulated private light sources (e.g. residential indoor lights spilling outdoors) can be estimated for each city or metropolitan area as a fractional contribution to the total with the help of basic urban statistics and direct sampling campaigns (Gokus et al, 2023; Aubé and Houle, 2023).

## 5. Conclusions

The variability of the optical properties of the atmosphere limits the minimum changes in anthropogenic light emissions that can be reliably detected by usual radiance sensing procedures. All measurement approaches are affected to a greater or lesser extent by this variability, depending on the particular combination of measurement geometry and state of the atmosphere.

If the only uncertainty is due to the fluctuations of the molecular and aerosol optical depths, the best performance for detecting small emission changes could be achieved by measuring the direct radiance of the sources from nearby vantage points, as long as the distance from the observer to the sources is kept small (~1km, for the combination of parameters analyzed in this work). At that particular distance, the minimum detectable changes in emissions are of order 1%, 2%, and 3% for the three variabilities of the aerosol optical depth analyzed in this work, $\sigma_{\tau_{0A}}(@550\text{ nm}) = [0.004, 0.01, 0.02]$, respectively. However, the performance of this method quickly worsens when the distance is increased, reaching the 10% at 15 km, 5 km, and 3 km, respectively. Satellite measurements allow to detect small changes as long as the airmass number is small (near-nadir observations), outperforming the other methods for distances larger than ~1 km, excepting for the distance interval in which the zenith scattering radiance sensing reaches its best results. Minimum detectable emission changes using satellite measurements are of order 1%, 2%, and 4%, for the variabilities $\sigma_{\tau_{0A}}(@550\text{ nm})$ assumed in this work. Finally, the minimum detectable emission changes using the scattered radiance approach depend on the distance to the sources in a non-monotonic way, reaching a minimum when the scattering and attenuation fluctuations compensate each other, situation in which this method tends to outperform satellite measurements, with detection limits below 1%. This happens at a distance of order of the photon mean free path (a few km, in our examples). The minimum detectable changes with this approach, for the parameters of our study, are in the intervals 0.7%–7%, 0.7%–20%, and 0.8%–40%, for the distance range 1–100 km and the three values of the aerosol optical depth variability $\sigma_{\tau_{0A}}$(@550 nm), respectively. For the examples analyzed in this work these limits depend more weakly on the variability of the molecular optical depth, $\sigma_{\tau_{0R}}$(@550 nm), which has been taken as an order of magnitude smaller than its aerosol counterpart.

In summary, the results of this paper provide two basic suggestions for practitioners in the light pollution monitoring community. First, detecting small inter-annual emission changes requires that the combined uncertainty of the measured annual mean radiances be kept sufficiently low. This implies, among other things, controlling the inter-annual variability of the mean atmospheric parameters within each yearly measurements dataset. This can be done by correcting each individual radiance reading for the instantaneous state of the atmosphere or by selecting a subset of measurements taken under similar atmospheric conditions each year. Sufficient number of measurements, to avoid undersampling, and good quality detectors are also required in order to reduce the uncertainty of the yearly mean radiance. Second, the method of choice for light pollution emissions monitoring that minimizes the uncertainty derived from a changing atmosphere depends on the distance of the observer to the sources. A careful analysis of both issues should be carried out when designing and deploying monitoring systems for evaluating the inter-annual evolution of the anthropogenic light emissions, in order to verify compliance with the public policies aimed at restoring the night.

## Acknowledgments

Almost no result in modern science stems from the work of a single author, and this paper is not an exception. Thanks are due to all members of the light pollution monitoring community for many fruitful and enriching discussions. The comments and suggestions of two Reviewers helped to improve significantly the text of this article. Any remaining error is the sole responsibility of the author.

## Funding / AI use

This study had no specific funding. AI tools have not been used in this work.

## References

Alarcon MR, Puig-Subirà M, Serra-Ricart M, Lemes-Perera S, Mallorquín M, López C. SG-WAS: A New Wireless Autonomous Night Sky Brightness Sensor. Sensors 2021;21:5590. https://doi.org/10.3390/s21165590

Aubé M, Simoneau A. New features to the night sky radiance model illumina: Hyperspectral support, improved obstacles and cloud reflection. J Quant Spectrosc Radiat Transf 2018;211:25-34, https://doi.org/10.1016/j.jqsrt.2018.02.033

Aubé M, Houle JP. Estimating lighting device inventories with the LANcube v2 multiangular radiometer. Int J Sustain Light 2023;25(1):10-23. https://doi.org/10.26607/ijsl.v25i1.131

Bará S. Anthropogenic disruption of the night sky darkness in urban and rural areas. R Soc Open Sci 2016;3:160541 https://doi.org/10.1098/rsos.160541

Bará S, Marco E, Ribas SJ, Garcia Gil M, Sánchez de Miguel A, Zamorano J. Direct assessment of the sensitivity drift of SQM sensors installed outdoors. Int J Sustain Light 2021;23:1–6. doi: https://doi.org/10.26607/ijsl.v23i1.109

Bará S, Rigueiro I, Lima RC. Monitoring transition: expected night sky brightness trends in different photometric bands. J Quant Spectrosc Radiat Transf 2019b;239:106644. https://doi.org/10.1016/j.jqsrt.2019.106644

Bará S, Rodríguez-Arós A, Pérez M, Tosar B, Lima RC, Sánchez de Miguel A, Zamorano J. Estimating the relative contribution of streetlights, vehicles, and residential lighting to the urban night sky brightness. Lighting Res. Technol. 2019;51:1092–1107. https://doi.org/10.1177/1477153518808337

Bará S, Tapia CE, Zamorano J. Absolute Radiometric Calibration of TESS-W and SQM Night Sky Brightness Sensors. Sensors 2019;19(6):1336. https://doi.org/10.3390/s19061336

Benavent-Oltra JA, Román R, Casquero-Vera JA, Pérez-Ramírez D, Lyamani H, Ortiz-Amezcua P, Bedoya-Velásquez AE, de Arruda Moreira G, Barreto Á, Lopatin A, Fuertes D, Herrera M, Torres B, Dubovik O, Guerrero-Rascado JL, Goloub P, Olmo-Reyes FJ, Alados-Arboledas L. Different strategies to retrieve aerosol properties at night-time with the GRASP algorithm. Atmos Chem Phys 2019;19:14149–14171. https://doi.org/10.5194/acp-19-14149-2019.

Bodhaine BA, Wood NB, Dutton EG, Slusser JR. On Rayleigh Optical Depth Calculations. J Atmos Oceanic Technol 1999;16:1854–1861. https://doi.org/10.1175/1520-0426(1999)016<1854:ORODC>2.0.CO;2

Cavazzani S, Ortolani S, Bertolo A, Binotto R, Fiorentin P, Carraro G, Zitelli V. Satellite measurements of artificial light at night: aerosol effects. Mon Notices Royal Astron Soc 2020;499(4):5075–5089. https://doi.org/10.1093/mnras/staa3157

Cinzano P. Night Sky Photometry with Sky Quality Meter. Internal Report No.9, v.1.4. Istituto di Scienza e Tecnologia dell'Inquinamento Luminoso (ISTIL), 2005. https://www.researchgate.net/publication/228399779_Night_Sky_Photometry_with_Sky_Quality_Meter

Cinzano P, Falchi F. The propagation of light pollution in the atmosphere. Mon Notices Royal Astron Soc 2012;427:3337–3357. https://doi.org/10.1111/j.1365-2966.2012.21884.x

Dobler G, Bianco FB, Sharma MS, Karpf A, Baur J, Ghandehari M, Wurtele J, Koonin SE. The Urban Observatory: A Multi-Modal Imaging Platform for the Study of Dynamics in Complex Urban Systems. Remote Sens 2021;13:1426. https://doi.org/10.3390/rs13081426

Dobler G, Ghandehari M, Koonin SE, Nazari R, Patrinos A, Sharma MS, Tafvizi A, Vo HT, Wurtele JS. Dynamics of the urban lightscape. Information Systems 2015;54:115-126. https://doi.org/10.1016/j.is.2015.06.002

Dobler G, Ghandehari M, Koonin SE, Sharma MS. A Hyperspectral Survey of New York City Lighting Technology. Sensors 2016;16(12):2047. https://doi.org/10.3390/s16122047

Dutton EG, Reddy P, Ryan S, DeLuisi JJ. Features and effects of aerosol optical depth observed at Mauna Loa, Hawaii: 1982–1992. J Geophys Res 1994;99(D4):8295–8306. https://doi.org/10.1029/93JD03520

Dubovik O, Holben B, Eck TF, Smirnov A, Kaufman YJ, King MD, et al. Variability of absorption and optical properties of key aerosol types observed in worldwide locations. J Atmos Sci 2002;59:590-608. https://doi.org/10.1175/1520-0469(2002)059<0590:VOAAOP>2.0.CO;2

Elvidge CD, Baugh K, Ghosh T, Zhizhin M, Hsu FC, Sparks T, Bazilian M, Sutton PC, Houngbedji K, Goldblatt R. Fifty years of nightly global low-light imaging satellite observations. Front. Remote Sens 2022;3:919937. https://doi.org/10.3389/frsen.2022.919937

Elvidge CD, Baugh K, Zhizhin M, Hsu FC, Ghosh T. VIIRS night-time lights. Int J Remote Sens 2017;38(21):5860-5879. https://doi.org/10.1080/01431161.2017.1342050

Espey BR. Empirical Modelling of Public Lighting Emission Functions. Remote Sens 2021;13(19):3827. https://doi.org/10.3390/rs13193827

Falchi F, Bará S. A linear systems approach to protect the night sky: implications for current and future regulations. R Soc Open Sci 2020;7:201501. https://doi.org/10.1098/rsos.201501

Falchi F, Cinzano P, Duriscoe D, Kyba CCM, Elvidge CD, Baugh K, Portnov BA, Rybnikova NA, Furgoni R. The new world atlas of artificial night sky brightness, Sci Adv 2016;2:e1600377. https://doi.org/10.1126/sciadv.1600377

Fernández AJ, Molero F, Salvador P, Revuelta A, Becerril-Valle M, Gómez-Moreno FJ, Artíñano B, Pujadas M. Aerosol optical, microphysical and radiative forcing properties during variable intensity African dust events in the Iberian Peninsula. Atmospheric Research 2017;196:129–141. http://doi.org/10.1016/j.atmosres.2017.06.019

Garstang RH. Night-sky brightness at observatories and sites. Pub Astr Soc Pac 1989;101:306-329. https://doi.org/10.1086/132436

Giles DM, Sinyuk A, Sorokin MG, Schafer JS, Smirnov A, Slutsker I, Eck TF, Holben BN, Lewis JR, Campbell JR, Welton EJ, Korkin SV, Lyapustin AI. Advancements in the Aerosol Robotic Network (AERONET) Version 3 database – automated near-real-time quality control algorithm with improved cloud screening for Sun photometer aerosol optical depth (AOD) measurements. Atmos Meas Tech 2019;12:169–209. https://doi.org/10.5194/amt-12-169-2019

Gokus A, Hänel A, Ruby A, Dröge-Rothaar A, Küchly B, Kyba CCM, Fischer D, et al. The Nachtlichter app: a citizen science tool for documenting outdoor light sources in public spaces. Int J Sustain Light 2023;25(1):24-59. https://doi.org/10.26607/ijsl.v25i1.133

Hänel A, Posch T, Ribas SJ, Aubé M, Duriscoe D, Jechow A, Kollath Z, Lolkema DE, Moore C, Schmidt N, Spoelstra H, Wuchterl G, Kyba CCM. Measuring night sky brightness: methods and challenges. J Quant Spectrosc Radiat Transf 2018;205:278–290. https://doi.org/10.1016/j.jqsrt.2017.09.008

Hess M, Koepke P, Schult I. Optical Properties of Aerosols and Clouds: The Software Package OPAC. Bull Amer Meteor Soc 1998;79(5):831-844. https://doi.org/10.1175/1520-0477(1998)079<0831:OPOAAC>2.0.CO;2

Horvath H, Alados Arboledas L, Olmo FJ, Jovanovic O, Gangl M, Kaller W, Sánchez C, Sauerzopf H, Seidl S. Optical characteristics of the aerosol in Spain and Austria and its effect on radiative forcing. J Geophys Res 2002;107(D19):4386. https://doi.org/10.1029/2001JD001472

Jones A, Noll S, Kausch W, Szyszka C, Kimeswenger S. An advanced scattered moonlight model for Cerro Paranal. A&A 2013;560:A91. https://doi.org/10.1051/0004-6361/201322433

Kasten F, Young AT. Revised optical air mass tables and approximation formula. Appl Opt 1989;28(22):4735-4738. https://doi.org/10.1364/AO.28.004735

Kieffer HH, Stone TC. The spectral irradiance of the Moon. The Astronomical Journal 2005;129:2887–2901. https://doi.org/10.1086/430185

Kocifaj M. Light-pollution model for cloudy and cloudless night skies with ground-based light sources Appl Opt 2007;46:3013-3022. https://doi.org/10.1364/AO.46.003013

Kocifaj M. Sky luminance/radiance model with multiple scattering effect. Solar Energy 2009;83(10):1914–1922. https://doi.org/10.1016/j.solener.2009.07.004

Kocifaj M. Retrieval of angular emission function from whole-city light sources using night-sky brightness measurements. Optica 2017;4:255-262. doi: 10.1364/OPTICA.4.000255

Kocifaj M. Multiple scattering contribution to the diffuse light of a night sky: A model which embraces all orders of scattering. J Quant Spectrosc Radiat Transf 2018;206:260-272. https://doi.org/10.1016/j.jqsrt.2017.11.020

Kocifaj M. Towards a comprehensive city emission function (CCEF). J Quant Spectrosc Radiat Transf 2018b;205:253-266. https://doi.org/10.1016/j.jqsrt.2017.10.006

Kocifaj M, Wallner S, Gueymard CA. Detailed assessment of multiple-scattering effects on night-sky brightness modeling in turbid environments: The impact of truncation and convergence errors. J Geophys Res: Atmospheres 2024;129:e2023JD039804. https://doi.org/10.1029/2023JD039804

Kokaly RF, Clark RN, Swayze GA, Livo KE, Hoefen TM, Pearson NC, et al. USGS Spectral Library Version 7: U.S. Geological Survey Data Series 1035; 2017. https://doi.org/10.3133/ds1035

Korras-Carraca MB, Hatzianastassiou N, Matsoukas C, Gkikas A, Papadimas CD. The regime of aerosol asymmetry parameter over Europe, the Mediterranean and the Middle East based on MODIS satellite data: evaluation against surface AERONET measurements. Atmos Chem Phys 2015;15:13113–13132. https://doi.org/10.5194/acp-15-13113-2015

Kyba CCM, Altıntas YO, Walker CE, Newhouse M. Citizen scientists report global rapid reductions in the visibility of stars from 2011 to 2022. Science 2023;379:265–268. https://doi.org/10.1126/science.abq7781

Kyba CCM, Garz S, Kuechly H, Sánchez de Miguel A, Zamorano J, Fischer J, Hölker F. High-Resolution Imagery of Earth at Night: New Sources, Opportunities and Challenges. Remote Sens 2015;7:1-23, https://doi.org/10.3390/rs70100001

Kyba CCM, Kuester T, Sánchez de Miguel A, Baugh K, Jechow A, Hölker F, Bennie J, Elvidge CD, Gaston KJ, Guanter L. Artificially lit surface of Earth at night increasing in radiance and extent. Sci Adv 2017;3(11):e1701528. https://doi.org/10.1126/sciadv.1701528

Kyba CCM, Wagner JM, Kuechly HU, Walker CE, Elvidge CD, Falchi F, Ruhtz T, Fischer J, Hölker F. Citizen Science Provides Valuable Data for Monitoring Global Night Sky Luminance. Sci Rep 2013;3:1835 https://doi.org/10.1038/srep01835

Li X, Elvidge C, Zhou Y, Cao C, Warner T. Remote sensing of night-time light. Int J Remote Sens 2017;38(21):5855-5859. https://doi.org/10.1080/01431161.2017.1351784

Masana E, Carrasco JM, Bará S, Ribas SJ. A multi-band map of the natural night sky brightness including Gaia and Hipparcos integrated starlight. Mon Notices Royal Astron Soc 2021;501:5443–5456. https://doi.org/10.1093/mnras/staa4005

McComiskey A, Schwartz SE, Schmid B, Guan H, Lewis ER, Ricchiazzi P, Ogren JA. Direct aerosol forcing: Calculation from observables and sensitivities to inputs. J Geophys Res 2008;113:D09202. https://doi.org/10.1029/2007JD009170

Paris, France 2024. Eclairage public. https://opendata.paris.fr/explore/dataset/eclairage-public/map/? (last accessed, 15th May 2024)

Pascual S, Zamorano J, González E, Tapia C, González R, García C, García L, Sánchez-Penim A, Izquierdo J, Corcho Ó, Sánchez de Miguel A, Gallego J. Citizen Science with the TESS Photometer Network. Astronomical Data Analysis Software and Systems XXX, ASP Conference Series, Vol. 532. J. E. Ruiz, F. Pierfederici, and P. Teuben, eds. 2021 Astronomical Society of the Pacific.

Posch T, Binder F, Puschnig J. Systematic measurements of the night sky brightness at 26 locations in Eastern Austria. J Quant Spectrosc Radiat Transf 2018;211:144–165. https://doi.org/10.1016/j.jqsrt.2018.03.010

Puschnig J, Näslund M, Schwope A, Wallner S. Correcting sky-quality-meter measurements for ageing effects using twilight as calibrator. Mon Notices Royal Astron Soc 2021;502:1095–1103. https://doi.org/10.1093/mnras/staa4019

Puschnig J, Schwope A, Posch T, Schwarz R. The night sky brightness at Potsdam-Babelsberg including overcast and moonlit conditions. J Quant Spectrosc Radiat Transf 2014;139:76–81. https://doi.org/10.1016/j.jqsrt.2013.12.011

Puschnig J, Wallner S, Schwope A, Näslund M. Long-term trends of light pollution assessed from SQM measurements and an empirical atmospheric model. Mon Notices Royal Astron Soc 2023;518(3):4449–4465. https://doi.org/10.1093/mnras/stac3003

Román R, Torres B, Fuertes D, Cachorro VE, Dubovik O, Toledano C, et al. Remote sensing of lunar aureole with a sky camera: Adding information in the nocturnal retrieval of aerosol properties with GRASP code. Remote Sens Environ 2017;196:238-252. https://doi.org/10.1016/j.rse.2017.05.013

Román MO, Wang Z, Sun Q, et al. NASA's Black Marble nighttime lights product suite, Remote Sens Environ 2018;210:113-143. https://doi.org/10.1016/j.rse.2018.03.017

Sánchez de Miguel A, Aubé M, Zamorano J, Kocifaj M, Roby J, Tapia C. Sky Quality Meter measurements in a colour-changing world. Mon Notices Royal Astron Soc 2017; 467(3): 2966-2979. https://doi.org/10.1093/mnras/stx145

Sánchez de Miguel A, Bennie J, Rosenfeld E, Dzurjak S, Gaston KJ. Environmental risks from artificial nighttime lighting widespread and increasing across Europe. Sci Adv 2022;8:eabl6891. https://doi.org/10.1126/sciadv.abl6891

Sánchez de Miguel A, Kyba CCM, Aubé M, Zamorano J, Cardiel N, Tapia C, Bennie J, Gaston KJ. Colour remote sensing of the impact of artificial light at night (I): The potential of the International Space Station and other DSLR-based platforms. Remote Sens Environ 2019;224:92–103. https://doi.org/10.1016/j.rse.2019.01.035

Sánchez de Miguel A, Zamorano J, Aubé M, Bennie J, Gallego J, Ocaña F, Pettit DR, Stefanov WL, Gaston KJ. Colour remote sensing of the impact of artificial light at night (II): Calibration of DSLR-based images from the International Space Station. Remote Sens Environ, 2021;264:112611. https://doi.org/10.1016/j.rse.2021.112611

Simoneau A, Aubé M, Leblanc J, Boucher R, Roby J, Lacharité F. Point spread functions for mapping artificial night sky luminance over large territories. Mon Notices Royal Astron Soc 2021;504(1):951–963. https://doi.org/10.1093/mnras/stab681

Stefanov WL, Evans CA, Runco SK, Wilkinson MJ, Higgins MD, Willis K. Astronaut Photography: Handheld Camera Imagery from Low Earth Orbit, in J.N. Pelton et al. (eds.), Handbook of Satellite Applications, Springer International Publishing Switzerland, 2017. https://doi.org/10.1007/978-3-319-23386-4_39

Teillet PM. Rayleigh optical depth comparisons from various sources. Appl Opt 1990;29:1897-1900. https://doi.org/10.1364/AO.29.001897

Thompson DW, Wallace JM. Regional climate impacts of the Northern Hemisphere annular mode. Science 2001;293(5527):85-89. https://doi.org/10.1126/science.1058958

van Loon H, Madden, RA.Interannual Variations of Mean Monthly Sea-Level Pressure in January. Journal of Climate and Applied Meteorology 1983;22(4):687–692. https://doi.org/10.1175/1520-0450(1983)022%3C0687:IVOMMS%3E2.0.CO;2

Vogel A, Alessa G, Scheele R, Weber L, Dubovik O, North P, Fiedler S. Uncertainty in aerosol optical depth from modern aerosol-climate models, reanalyses, and satellite products. J Geophys Res: Atmospheres 2022;127:e2021JD035483. https://doi.org/10.1029/2021JD035483

Wallace JM, Thompson DWJ. Annular Modes and Climate Prediction. Physics Today 2002;55(2):28–33. https://doi.org/10.1063/1.1461325

Wallner S, Kocifaj M. Impacts of surface albedo variations on the night sky brightness – A numerical and experimental analysis. J Quant Spectrosc Radiat Transf 2019;239:106648. https://doi.org/10.1016/j.jqsrt.2019.106648

Wallner S, Puschnig J, Stidl S. The reliability of satellite-based lighttrends for dark sky areas in Austria. J Quant Spectrosc Radiat Transf 2023:311:108774. https://doi.org/10.1016/j.jqsrt.2023.108774

Winkler H. A revised simplified scattering model for the moonlit sky brightness profile based on photometry at SAAO. Mon Notices Royal Astron Soc 2022;514(1):208–226. https://doi.org/10.1093/mnras/stac1387

Young, AT. On the Rayleigh-Scattering Optical Depth of the Atmosphere. J Appl Meteor Climatol 1981;20:328–330. https://doi.org/10.1175/1520-0450(1981)020<0328:OTRSOD>2.0.CO;2

Zamorano J, Bará S, Barco M, García C, Caballero AL. Controlling the artificial radiance of the night sky: the Añora urban laboratory. J Quant Spectrosc Radiat Transf 2023;296:108454. https://doi.org/10.1016/j.jqsrt.2022.108454

Zamorano J, García C, González R, Tapia C, Sánchez de Miguel A, Pascual S, Gallego J, González E, Picazo P, Izquierdo J, et al. STARS4ALL Night Sky Brightness Photometer. Int J Sustain Light 2017;35:49–54. https://doi.org/10.26607/ijsl.v18i0.21